\documentclass[a4paper,fleqn,usenatbib]{mnras}
\usepackage{dcolumn}
\usepackage{graphicx}
\usepackage{url}
\usepackage{color}
\usepackage[T1]{fontenc}
\usepackage{ae,aecompl}
\usepackage{pdflscape}

\newcommand{\cm}{cm$^{-1}$}

\newcommand{\ai}{\textit{ab initio}}

\newcommand{\ea}{\textit{et al.}}

\newcommand{\eqref}[1]{(\ref{#1})}

\newcommand{\hp}{H$_3^+$}
\newcommand{\hdp}{H$_2$D$^+$}
\newcommand{\dhp}{D$_2$H$^+$}

\title[ExoMol XX: Line list for H$_3^+$]{ExoMol molecular line lists  XX: 
a comprehensive line list for H$_3^+$}

\author[I.I. Mizus et al.]{Irina I. Mizus$^{1}$, Alexander Alijah$^2$, Nikolai F. Zobov$^1$,  
\newauthor Lorenzo Lodi$^3$, Aleksandra A. Kyuberis$^{1}$, Sergei N. Yurchenko$^{3}$, 
\newauthor  Jonathan Tennyson$^{3}$\thanks{Email: 
j.tennyson@ucl.ac.uk} and Oleg L. Polyansky$^{1,3}$,
\\
$^{1}$Institute of Applied Physics, Russian Academy of Sciences, Ulyanov Street 
46, Nizhny Novgorod 603950, Russia\\
$^{2}$Groupe de Spectrom\'etrie Mol\'eculaire et Atmosph\'erique,
GSMA, UMR CNRS 7331, Universit\'e de Reims Champagne-Ardenne, France\\
$^{3}$Department of Physics and Astronomy, University College London, London 
WC1E 6BT, UK}
\date{Accepted XXXX. Received XXXX; in original form XXXX}
\pubyear{2016}

\begin{document}
\label{firstpage}
\pagerange{\pageref{firstpage}--\pageref{lastpage}} 

\maketitle

\begin{abstract}
  H$_3^+$ is a ubiquitous and important astronomical species whose
  spectrum has been observed in the interstellar medium, planets and
  tentatively in the remnants of supernova SN1897a. Its role as a cooler is important
  for gas giant planets and exoplanets, and possibly the early
  Universe.    All this makes the spectral properties, cooling
  function and partition function of H$_3^+$ key parameters for
  astronomical models and analysis. A new high-accuracy, very
  extensive line list for H$_3^+$ called MiZATeP was computed as 
  part of the ExoMol project alongside a temperature-dependent cooling
  function and partition function as well as lifetimes for 
  excited states. These data are made available in electronic form as
  supplementary data to this article and at \url{www.exomol.com}.

\end{abstract}


\section{Introduction}

The atomic composition of the Universe is dominated by hydrogen which
means that H$_3^+$, as the stable ionic form of molecular hydrogen, is
thought to be important in many diverse astronomical environments
where it plays a variety of roles \citep{MO00,oka:2006}. So far
H$_3^+$ has been observed in the atmospheres of the solar system gas
giants \citep{jt80,jt127,gj93,jt155}, dense molecular clouds
\citep{g96,mccall:1999}, the diffuse interstellar medium
\citep{mg98,mh02} and external galaxies
\cite{06GeGoUs.H3+,15GeMaOk.H3+}, and more tentatively in the remnants
of supernova SN1897a \citep{jt110}. Observations of H$_3^+$ provide a
powerful tool for studying the Galactic centre
\citep{02GoMcGe.H3+,05OkGeGo.H3+,08GoUsNa.H3+}, where it has been
shown that lifetime effects in H$_3^+$ lead to populating
long-lived meta-stable states.  A similar mechanism is also important in
laboratory studies of H$_3^+$ \citep{jt306,jt340}.  So far searches
for H$_3^+$ in the atmosphere of hot Jupiter exoplanets have proved
negative \citep{06ShGaMo.H3+}, while the claimed detection of H$_3^+$
emission in a protoplanetary disk \citep{02BrRexx.H3+} was negated by
\citet{05GoGeMc.H3+}.

H$_3^+$, which is rapidly formed from the collision of molecular
hydrogen and its ion (H$_2^+$), has long been thought to be the
initiator of much of interstellar gas-phase chemistry
\citep{w73,hk73,jt157,13Oka.H3+,15Millar}.  It provides a unique means
to monitor cosmic-ray ionization rates in the interstellar medium
\citep{mccall:2003b,12InMc}. Cooling by H$_3^+$ is thought to be
important for the stability of atmospheres of giant extrasolar planets
orbiting close to their stars \citep{kam07,15KhShLa.H3+} and possibly
in primordial gas \citep{06GlSa}. Cooling is one of a number of
functions performed by H$_3^+$ in the ionospheres of solar system gas
giants \citep{jt258} where observations of H$_3^+$ have proved
important for monitoring ionospheric activity
\citep{jt167,jt192,jt201,jt258,08StMiMe.H3+,08StMiLy.H3+} and have,
for example, been used to determine wind speeds \citep{99ReAcSt.H3+}.
Elsewhere H$_3^+$ is probably a key component of cool stars with low
metallicity; for example it has been shown to play a crucial role in
the chemical evolution of cool white dwarfs \citep{brl97}.

H$_3^+$ has no known electronic spectrum and its \lq forbidden' pure rotational
spectrum, although possibly observable \citep{86PaOk.H3+,jt72}, has yet to
be detected. This leaves its vibration-rotation spectrum as the means by
which all spectroscopic studies are made. The laboratory spectroscopic
data for H$_3^+$ was recently collected and reviewed by \citet{13FuSzMa.H3+}
as part of their MARVEL, measured active rotational-vibrational energy levels \citep{jt412,12FuCs.method},
study of the system. This work replaced an earlier compilation and 
evaluation of the laboratory data by \citet{lindsay:2001a}. 
\citet{13FuSzMa.H3+} provide a set of empirical energy levels
for H$_3^+$ which we use below.

\citet{jt107}
presented a line list of 699 astronomically-important H$_3^+$ lines based on laboratory
transition frequencies and {\it ab initio} transition intensities.
The work was supplemented by \citet{jt181} (NMT below) who computed
a much more extensive  H$_3^+$ 
line list composed of about $3\times10^6$ lines. 
These calculations were based on the use of an
empirically-determined potential energy surface (PES) \citep{jt175}
and an {\it ab initio} dipole moment surface (DMS) \citep{lf92}.
The quality of this line list was determined first of all by the 
high accuracy of the fitted PES used for the calculation of the wavefunctions,
leading to a standard deviation with respect to the experimental
energy levels of only 0.009 \cm.
One of the reasons of this accuracy was the simultaneous fit of all \hp\ 
isotopologues, based on the accurate \ai\ determination of both symmetric and
asymmetric  adiabatic surfaces of \hdp\ and \dhp\ \citep{jt166}. 
Note that only states with energies up to 
15\,000~cm$^{-1}$ were considered in these studies; indeed, NMT regarded
their results for states above the barrier to linearity at about 10\,000~\cm\
as highly uncertain as at that time there was no available spectroscopic data
for H$_3^+$ probing this region. \citet{jt169}  provided a high-temperature
partition function for H$_3^+$ which was significantly larger at high-$T$
than some previous functions used by astronomers; they showed that such
values relied on considering all the levels up to the dissociation limit
of H$_3^+$ at about 35\,000 \cm. NMT also provided the first H$_3^+$
cooling function, which was refined in subsequent studies \citep{jt489,jt551}
also based on the NMT line list.

The NMT line list has been widely used for astronomical and other
studies.  For example the use of the NMT line list was instrumental in
assignment and reassignment of numerous experimentally observed lines
by \citet{jt193}.  It has also been shown to be very accurate for
spectroscopic intensity predictions \citep{jt512,jt587}, perhaps 
surprisingly so. However, improved theoretical
modelling of the spectroscopy of H$_3^+$, discussed below, implies that
we are now in position to compute a line list which is both more
accurate and more complete, as well as being able to rectify other known issues
with the NMT list. NMT performed nuclear motion calculations in Jacobi coordinates
and, as a consequence, their wavefunctions did not possess the full symmetry
of the system. This symmetry is important for determining whether a
state is ortho or para and hence whether its nuclear spin statistical
weight is 4 or 2. NMT assigned symmetry by hand to a few levels but
the vast majority were simply given the average statistical weight of
$\frac{8}{3}$. The use of lower symmetry meant that many of the
Einstein A coefficients computed should actually have been zero by
symmetry. Because of this and because their line list was very large
by contemporary standards, NMT removed all very weak transitions from
their line list. This had the unintended consequence of removing those
transitions which allow some long-lived meta-stable states of H$_3^+$
to decay by photon emission, which in turn limits the use of the NMT data for
modelling population trapping in H$_3^+$ and, by extension, for
constructing a reliable low-temperature cooling function.  We note
that the more recent line list for H$_2$D$^+$ computed by
\citet{jt478} does not suffer from these problems.

The present work provides
a new line list for H$_3^+$. Unlike NMT, the model used here is essentially
{\it ab initio}. H$_3^+$ is a two-electron system and is
a benchmark for developments in high accuracy {\it ab
initio} quantum chemical methods
\citep{Rohse1994,Cencek1998,jt236,03ScAlHia.H3+,03ScAlHib.H3+,kutzelnigg:2006,ptf09,jt526,jt566}.
Of particular note here is the non-adiabatic model developed by \citet{jt236} and the
ultra-high accuracy {\it ab initio} PES of \citet{jt526}. Use of these
were found to give frequency predictions of outstanding accuracy \citep{jt512}.
Theory has always played an important part in the astronomical spectroscopy
of H$_3^+$ since, as yet, there is only a single \citep{98McWa}  absolute laboratory measurements of H$_3^+$
line intensities. However, empirical tests of predicted intensities have also
been provided by experiments measuring intensity ratios for transitions
with widely differing wavelengths and intensities \citep{jt289,jt413,jt587}.
The most stringent test was provided by the visible-wavelength 
measurements of \citet{jt587} which showed that their DMS, used here,
predicted the observed intensities in a very satisfactory manner.

This new H$_3^+$ line list, which we call MiZATeP, is computed as part
of the ExoMol project \citep{jt528} which has provided a large number
of molecular line lists for exoplanet and other atmospheres
\citep{jt631}. The line lists produced by ExoMol to date are summarised
in Table~\ref{tab:exomoldata}; in addition, the BT2 H$_2$$^{16}$O
\cite{jt378} and BYTe NH$_3$ \cite{jt500} pre-dated the start of the
project.  H$_3^+$ is first  the molecular ion studied as part of the ExoMol
project, although line lists for H$_2$D$+$ \cite{jt478}, HeH$^+$
\cite{jt347}, HD$^+$ \cite{jt506} and LiH$^+$ \cite{jt506} were
computed previously.

\begin{table*}
\centering
\caption{Datasets created by the ExoMol project and included in the ExoMol
database.}
\label{tab:exomoldata}
\begin{tabular}{llcrrrll}
\hline\hline
Paper&Molecule&$N_{\rm iso}$&$T_{\rm max}$&$N_{elec}$&$N_{\rm lines}$
$^a$&DSName&Reference\\
\hline
I&BeH&1&2000 &1&16~400&Yadin& \citet{jt529}\\
I&MgH&3 &2000 &1&10~354&Yadin&\citet{jt529}\\
I&CaH&1 &2000 &1&15~278&Yadin& \citet{jt529}\\
II&SiO&5&9000&1& 254~675&EJBT&\citet{jt563}\\
III&HCN/HNC&2$^a$&4000&1&399~000~000&Harris& \citet{jt570}\\
IV&CH$_4$&1&1500&1&9~819~605~160&10to10&\citet{jt564}
\\
V&NaCl&2&3000&1& 702~271 &Barton&\citet{jt583}\\
V&KCl&4&3000&1& 1~326~765  &Barton& \citet{jt583}\\
VI&PN&2&5000&1&142~512&YYLT& \citet{jt590}\\
VII&PH$_3$&1&1500&1&16~803~703~395&SAlTY&  \citet{jt592}\\
VIII&H$_2$CO&1&1500&1&10~000~000~000&AYTY&  \citet{jt597}\\
IX&AlO&4&8000&3&4~945~580&ATP& \citet{jt598}\\
X&NaH&2&7000&2&79~898&Rivlin& \citet{jt605}\\
XI&HNO$_3$&1&500&1&6~722~136~109&AlJS& \citet{jt614}\\
XII&CS&8&3000&1&548~312&JnK& \citet{jt615}\\
XIII&CaO&1&5000&5&21~279~299&VBATHY& \citet{jt618}\\
XIV&SO$_2$&1&2000&1&1~300~000~000&ExoAmes& \citet{jt635}\\
XV&H$_2$O$_2$&1&1250&1&20~000~000~000&APTY& \citet{jt638}\\
XVI&H$_2$S&1&2000&1&115~530~373&AYT2& \citet{jt640}\\
XVII&SO$_3$&1&800&1&21~000~000~000&UYT2& \citet{jt635}\\
XVIII&VO&1&5000&13&277~131~624&VOMYT& \citet{jt644}\\
XIX&H$_2$O&2$^b$&3000&1&519~461~789&HotWat78& \citet{jt665}\\
XX&H$_3^+$&1&5000&1&127~542~657&MiZATeP&This work\\
XXI&NO&6&5000&1&2~281~042&NOname& \citet{jt686}\\
XXII&H$_2$O&1&3000&1&12~000~000~000&Pokazatel& \citet{jtpoz}\\
\hline
\hline
\end{tabular}

$N_{\rm iso}$ Number of isotopologues considered;\\
$T_{\rm max}$ Maximum temperature for which the line list is complete;\\
$N_{elec}$ Number of electronic states considered;\\
$N_{\rm lines}$  Number of lines: value is for the main isotope.\\
$^a$ A line list for H$^{13}$CN/HN$^{13}$C due to Harris \ea\ \citet{jt447} 
is
also available.\\
$^b$ HotWat78 are line lists for H$_2$$^{18}$O and  H$_2$$^{17}$O in the
style of the BT2  H$_2$$^{16}$O \citep{jt378} and VTT HDO \citep{jt469}
line lists. Pokazatel, number XXII, is an extended
 H$_2$$^{16}$O line list.
\end{table*}

\section{Method}\label{method}
Nuclear motion calculations used the highly accurate global \emph{ab initio} PES 
presented by \citet{jt526} and the related DMS given by \citet{jt587}. The DMS 
is expressed in the 7-parameter form of \citet{lf92} which was found to best
reproduce the observations.
The calculations were based on the {\sc DVR3D} program suite \citep{jt338}
and were performed for two different choices of the basis set and were 
augmented by a third set of calculations for labelling purposes 
performed using a separate program by \citet{WOL88:371}.

The bulk of the calculations were performed in Jacobi coordinates and
used the \citet{jt236} model to allow for non-adiabatic effects.
Discrete variable representation (DVR) grids were based on spherical
oscillator functions \citep{jt23} for both the atom -- diatom
coordinate and diatomic \citep{jt14} coordinate, and (associated) Legendre
functions  for the angular coordinate. The grids contained
$60$, $58$, and $68$ points for these coordinates, respectively.  The
final diagonalized matrices for the vibrational problem had a dimension
of $20\,000$. Further increases of these parameters do not lead to
significant changes in the resulting energies.  These calculations
used spherical oscillators with parameters $\alpha = 0.0$ and
$\omega_e = 0.07$ atomic units for both radial coordinates.
Non-adiabatic effects were taken into account by using different
values for the vibrational and the rotational masses in the kinetic
energy operator; the vibrational mass was taken to be equal to
1.007537~D$_a$ -- an intermediate value between nuclear and atomic
masses suggested by \citet{96Moss} on the basis of calculations on
H$_2^+$ isotopologues. The proton (nuclear) mass was used for the
rotational mass.  These calculations yielded energy levels
up to at least 25\,000~cm$^{-1}$ for $J$ values up to 25.
The model used for the calculation has been shown to give an accuracy
of about 0.1 \cm \citep{jt512,jt526} for all experimentally observed
energy levels. The highest energy level lies at about 17~000~\cm. As the PES is \ai\ 
we hope that this accuracy extrapolates well to all the energies
used in the presented line list.

These {\sc DVR3D} calculations with big basis were supplemented by
second set of smaller calculations which used $31$, $31$, and $50$
grid points for two radial and an angular coordinates, respectively,
and with the final vibrational Hamiltonians dimensions equal to
$3000$. The calculations were performed up to at least
35~000~cm$^{-1}$ and for $J$ values 0 -- 40. Note that the highest
bound rotational state for H$_3^+$ is predicted to have $J=42$
\citep{jt64,13JaCa}.  These calculations used Morse-like oscillators
\citep{jt14} with parameters $r_e=3.1$, $D_e=0.1$, and $\omega_e =
0.006$ in atomic units for the diatomic radial coordinate and
spherical oscillators with parameters $\alpha=0.0$, and
$\omega_e=0.016$ atomic units for the scattering coordinate. Only
nuclear masses were used for these calculations.  This set of
calculations was performed to achieve better convergence for the
partition function and to provide completeness for the final line list
by adding transitions to energy states with $J$ values larger than 25.
Similar, more approximate treatments of the higher-lying states have
been used successfully for other ExoMol line lists and partition sums
\citep{jt571,jt635,jt641}.

Although it is possible to obtain full symmetrization of the {\sc
DVR3D} wavefunctions computed in Jacobi coordinates \citep{jt358},
here we achieved this goal by performing a third set of nuclear motion
calculations using the hyperspherical harmonics code of
\citet{WOL88:371}.  The hyperspherical coordinates as defined by
\citet{WHI68:1103} and modified by \citet{JOH83:1916} are the three
internal coordinates consisting of the hyperradius, $\rho$, and the
two hyperangles $\theta$ and $\phi$, and the three Euler angles
$\alpha$, $\beta$ and $\gamma$.  The symbol $\Omega$ is used to
collect the five angles, $\Omega= (\theta, \phi, \alpha, \beta,
\gamma)$.  In these coordinates, the Hamiltonian is written as
\begin{equation}\label{HC1}
H(\rho, \Omega) =
- \frac{\hbar^2}{2 \mu} \left[ \frac{1}{\rho^5} \frac{\partial}{\partial \rho}
\rho^5 \frac{\partial}{\partial \rho} + \frac{\Lambda^2(\Omega)}{\rho^5} \right] + V(\rho, \theta, \phi),
\end{equation}
where $\mu = \sqrt{m_1 m_2 m_3/(m_1+m_2+m_3)}$ is the three-particle
reduced mass and $\Lambda^2(\Omega)$ the grand angular momentum
operator.  Its eigenfunctions are the hyperspherical harmonics,
$\theta^{J\Gamma}_{\alpha}(\Omega)$.  As shown by \citet{WOL93:2695},
they can be symmetrized easily in the three-particle permutation
inversion group $S_3 \times I$. The labels are then the total angular
momentum $J$, the symmetry index $\Gamma$, and $\alpha$, a counting
index.  To solve the rovibrational Schr\"odinger equation
corresponding to Hamiltonian~(\ref{HC1}), the rovibrational wave
function is expanded in terms of symmetrized hyperspherical harmonics
\begin{equation}\label{HC2}
\Psi^{J \Gamma}_n(\rho, \Omega) = \sum_{\alpha} \theta^{J\Gamma}_{\alpha}(\Omega) \frac{P^{J \Gamma}_{\alpha, n}(\rho)}{\rho^{5/2}}.
\end{equation}
This yields a system of coupled equations in the hyperradius which is
integrated numerically. As the expansion converges only slowly, a
contracted basis of symmetrized hyperspherical harmonics is used. The
contraction coefficients are the lowest eigenvectors obtained from
diagonalization of the potential energy matrix, $U(\rho)$, with matrix
elements $U_{\alpha, \alpha'}(\rho)= \langle
\theta^{J\Gamma}_{\alpha}(\Omega) | V(\rho, \theta, \phi) |
\theta^{J\Gamma}_{\alpha'}(\Omega) \rangle_{\Omega}$ in the spherical
harmonics basis at a $\rho$ value that corresponds to the minimum of
the potential, $\rho=2.21 \, a_0$.  The procedure is fully described
by \citet{SCH03:175}.  Typically, about 1000 primitive hyperspherical
harmoncis are contracted to 300 basis functions, hence a system of 300
coupled equations is integrated. For each value of $J$, there are in
general six irreducible representations: $A_1'$, $A_2'$, $E'$,
$A_1''$, $A_2''$, $E''$.  Prime representations have even parity,
while double prime representations have odd parity. Hence for $J=0$
there are only three even parity representations.

For the production runs the code was modified so
that for each $\Gamma$ and $J$ the number of basis functions is
determined automatically so that, for a given symmetry, only the value
of $J$ needs to be set in the input. Numerical integration is done
within $0.7 \, a_0 \le \rho \le 6.2 \, a_0$, with a step size of
$\Delta \rho = 0.01 \, a_0$. The energy range of the desired
eigenvalues is split into six parts, and six separate jobs are run to
compute the eigenvalues within their respective energy intervals. In
the present implementation of the code no eigenfunctions are obtained,
which would be needed for the intensity calculations. The {\sc DVR3D}
code was used for this purpose.  On the other hand, the hyperspherical
code fully exploits permutational symmetry, thus allowing the identification
of degenerate states; such degenerate states appear in unsymmetrized 
{\sc DVR3D} calculations as a pair of $A_1$, $A_2$ states with very 
similar energy.

The hyperspherical harmonic calculations were used to provide full
symmetry labels for states obtained using {\sc DVR3D}. This labelling
procedure was performed for the first set of high accuracy calculations and
was limited to $J$ values up to $20$ only.  The (quasi-) degenerate
even and odd pairs of {\sc DVR3D} levels which correspond to
degenerate $f$-symmetry levels were identified. These levels are para
and have a nuclear-spin degeneracy factor of $2$. The degeneracy
factor for the $A_2$-type levels (the unmatched odd levels) is $4$.
Unmatched even levels are of $A_1$-type which have zero statistical
weight; these levels were discarded.

For higher $J$ we used the procedure suggested by
\citet{jt169} to set the nuclear spin degeneracy factor for
transitions between energy levels with $J$ values 21 -- 40 in our
final line list. This method avoids explicit labelling by using the
high-temperature approximation of ascribing a degeneracy factor
equal to $\frac83$ to odd levels, and equal to $0$ to even ones.
This removes the need to decide if a given pair of levels should be
degenerate and therefore of $E$-type, which becomes increasingly
difficult as the calculations are less well converged \citep{jt133}.
Given the small contribution of these high $J$ states, this procedure
introduces negligible error in the results given below.

\section{Line list calculations}\label{linelist}
A comprehensive line list was calculated for transition frequencies up to
25\,000~cm$^{-1}$. This line list comes in the form of a states
file, which stores energy levels and other state-specific information,
and a transitions file. Where available levels from the MARVEL analysis
 \citep{13FuSzMa.H3+} were used to replace our calculated values to ensure
the highest possible accuracy. 


This new H$_3^+$ line list, which we call MiZATeP, contains
transitions between energy states with $J$ values 0 -- 37 and energies
0 -- 42\,000~cm$^{-1}$ and consists of $127\,542\,657$ lines with an
accuracy close to the spectroscopic one; the $158\,721$ states
considered have rotational quantum numbers up to $J =37$. On the basis
of the calculated energy levels and taking into account their
statistical weights we also compute accurate partition and cooling
functions, which, we believe, are appropriate for temperatures
up to $5000$ K. The line list should also be valid up to this temperature.
The line list is presented in the updated ExoMol format
\citep{jt631}; extracts from the states and transitions files are
presented in tables \ref{tab:states} and \ref{tab:trans},
respectively.

The energies used in the states file are a mixture: (1) MARVEL energies
\citep{13FuSzMa.H3+} were used where available; (2) for $J \leq 25$
the high-quality results from the first set of nuclear motion
calculations were used; (3) for $J = 26 - 37$ the results of the
second set of calculation, performed with the smaller 
basis set, were used.
Levels with $J = 25$ required separate consideration, because
transitions between states with $J = 24$ and $J = 25$ (and
$25\,\longleftrightarrow\,25$) are a part of our accurate results,
whereas transitions between states with $J = 25$ and $J = 26$ were
treated using the results of the calculations with the small basis
set. Thus, the states file contains two sets of energy levels with $J = 25$:
the accurate ones and the ones obtained within the
small basis set. All energy
values are given relative to the same high-accurate value of 
ground state energy.
Whenever possible the states have been assigned quantum numbers following
the convention of \citet{84Watson.H3+}. In particular, the energy of a rovibrational state
can be expanded as, according to \citet{84Watson.H3+},
\begin{equation}
E(J,G) = T_0 + B J(J+1) + (C-B) G^2 + \cdots
\end{equation}
where $G=|k-\ell_2|$ and $\ell_2$ is the vibrational angular momentum.
Since, by convention, $C < B$ holds for the rotational constants,
the rotational energy increases, for a given vibrational state and $J$,
with decreasing $G$. It is reasonable to assume that
the states with infinite lifetime (see below) belong to the vibrational ground state
and have the largest values of $G$, i.e., $G \equiv K = J$ and $G \equiv K = J-1$.
We then determine the symmetry of these states, which is
$A_1/A_2$ for $G =0, 3, 6, \cdots$ (with just one state for $G=0$)
and $E$ for $G=1, 4, 7, \cdots$ and $G=2, 5, 8, \cdots$.
Prime and double prime lables are according to even or odd parity, respectively, 
of $G+v_2$. To assign the states in question, we simply pick,
of the eigenvalues computed in full symmetry with the hyperspherical
harmonics code,  the lowest one with the appropriate symmetry. 
This procedure works, because the lowest rotational levels of the next higher
vibrational states, $(0,1^1)$ and $(1,0^0)$, are well separated in energy.
The tag $-1$ is used for states for which no approximate quantum number assignments 
 are made.

\begin{figure}
\includegraphics[width=\columnwidth]{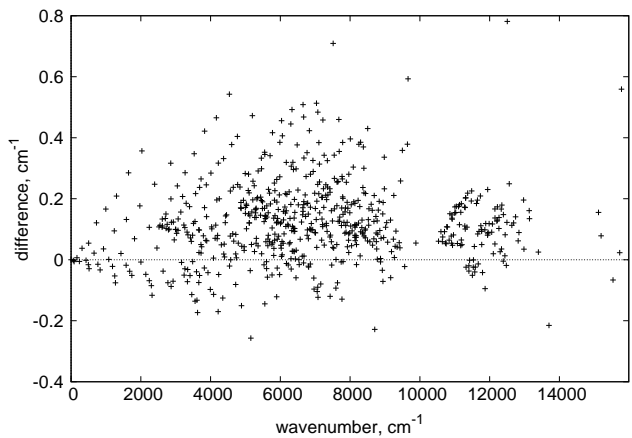}
\caption{Comparison of our energy levels calculations results with experimental energy values obtained during MARVEL analysis \citep{13FuSzMa.H3+}.}
\label{obs-calc}
\end{figure}

Figure \ref{obs-calc} shows the result of a comparison of our calculated energy values with almost all available
MARVEL energies of states with $J$ values up to 12. Standard deviation between theory and experiment here 
is about 0.18 \cm. 

While calculating the final version of our line list, lifetimes, partition and
cooling function values, it is only necessary to consider states
with odd vibrational symmetry \citep{jt338} in the {\sc DVR3D} calculation; these
states include both $E$ (one component) and  $A_2$ symmetry states. $A_1$
states have even symmetry and need not be considered. 

Statistical weights were assigned to almost all states with $J \leq 20$
 and energies up to 25\,000~cm$^{-1}$ through our labelling procedure. 
These weights are equal to $2$ for $E$ states 
and  $4$ for $A_2$ states. States outside this range
are given the average statistical weight of  $g_{\rm ns}=\frac{8}{3}$. To
retain compatibility with the ExoMol format \citep{jt631} for these
states, the product
$g_{\rm ns} \times (2J+1)$, which gives the total degeneracy of each level,
$g$, was rounded to the nearest integer.

The {\sc DVR3D} program suite for triatomic molecules 
does not, when using Jacobi coordinates, take into account 
the symmetry of the system when some of the nuclei are
identical, such as in the case of H$_3^+$.
As a consequence {\sc DVR3D} also calculates transitions which are 
forbidden by the exact H$_3^+$ selection rules, thus
producing in the resulting line list many very weak transitions 
which should actually have zero intensity.  
We systematically deleted such unwanted transitions from
our final line list, but there remains a possibility that there are
some allowed but very weak transitions that also got mistakenly
deleted due to errors in the labelling procedure.

Intensity calculations were based on the DMS by \citet{jt587}, which
has been expanded to an energy region up to 30\,000~cm$^{-1}$ to cover
all the frequency range needed for our goals.  Table \ref{tab_DMS}
presents a comparison of the
calculated Einstein $B$ coefficients obtained using the
DMS of \citet{jt587}, the results of NMT and a new calculation using the
DMS of \citet{Rohse1994}, with the experimental data from Table I of
\citet{jt587}. The standard
deviation of the ratio of experimental to calculated values is 22\%.
The comparison between our calculations with the two DMS
suggests that the main source of sensitivity in the intensity calculations is 
the DMS employed and not the wavefunctions and the underlying PES. The DMS
of \citet{jt587} covers a frequency range about twice as large as the one
considered by NMT \citep{jt181}, and is only slightly worse in energy
region up to 15\,000~cm$^{-1}$ -- the difference is about 4.5\% for
the same set of experimental data.



\begin{table}
\begin{center}
\caption{Comparison of the calculated Einstein's coefficients, $B$, obtained here 
with the DMS of \citet{jt587} ($B_{\rm calc}$), with the experimental data ($B_{\rm exp}$, \citet{jt587}), 
and also the results of NMT ($B_{\rm NMT}$, \citet{jt169}) and calculations made with DMS of \citep{Rohse1994} on the basis 
of the PES from \citet{jt526} ($B_{\rm R}$). $B$ values are measured in 
units $10^{18}$~cm$^3$~J$^{-1}$~s$^{-2}$. The transition 
frequencies, $\nu$, are taken from the larger DVR3D calculations, see text.}
\small
{\setlength{\tabcolsep}{1.5pt}
\begin{tabular}{cccccccc}
\hline\hline
$\nu$ (cm$^{-1}$) & $B_{\rm exp}$ & $B_{\rm NMT}$ & $B_{\rm R}$ & 
$B_{\rm calc}$ & $\frac{B_{\rm exp}}{B_{\rm NMT}}$ & $\frac{B_{\rm exp}}{B_{\rm R}}$ & $\frac{B_{\rm exp}}{B_{\rm calc}}$\\[4pt]
\hline
 $7144.005$    &            & $1550.00$ & 1554.31 & $1565.21$ &        &     &        \\
 $10\,752.085$ & $72.6(16)$ & $60.53$   & 60.297 & $55.864$ & $1.20$ & 1.20 & $1.30$ \\
 $10\,798.626$ & $26.5(30)$ & $32.94$   & 32.964 & $32.327$ & $0.80$ & 0.80 & $0.82$ \\
 $10\,831.526$ & $112(16)$  & $93.98$   & 94.078 & $96.168$ & $1.19$ & 1.19 & $1.16$ \\
 $12\,373.310$ & $4.3(10)$  & $4.040$   & 4.0141  & $4.5837$  & $1.06$ & 1.07 & $0.94$ \\
 $12\,381.054$ & $4.1(10)$  & $4.097$   & 4.0230  & $4.2277$  & $1.00$ & 1.02 & $0.97$ \\
 $12\,413.273$ & $4.6(12)$  & $3.734$   & 3.7896  & $3.7869$  & $1.23$ & 1.21 & $1.21$ \\
 $12\,588.962$ & $1.10(38)$ & $0.8589$  & 0.8599  & $0.7590$  & $1.28$ & 1.28 & $1.45$ \\
 $12\,620.082$ & $6.3(12)$  & $4.858$   & 4.7392  & $4.4116$  & $1.30$ & 1.33 & $1.43$ \\
 $12\,678.540$ & $8.6(17)$  & $8.006$   & 8.0987  & $8.4727$  & $1.07$ & 1.06 & $1.02$ \\
 $13\,332.856^*$ & $4.0(13)$  & $2.045$   & $2.0550$  & $1.7871$  & $1.96$ & $1.95$ & $2.24$ \\
 $13\,638.464$ & $3.9(15)$  & $4.137$   & 4.0570  & $3.6346$  & $0.94$ & 0.96 & $1.07$ \\
 $15\,058.522$ & $1.53(33)$ & $1.6189$  & 1.5916  & $1.3920$  & $0.95$ & 0.96 & $1.10$ \\
 $15\,130.399$ & $0.72(16)$ & $0.7120$  & 0.6979  & $0.8488$  & $1.01$ & 1.03 & $0.85$ \\
 $15\,450.172$ & $0.75(10)$ & $0.7747$  & 0.7716  & $0.7593$  & $0.97$ & 0.97 & $0.99$ \\
 $15\,643.023$ & $1.11(15)$ & $1.0125$  & 1.0103  & $1.0079$  & $1.10$ & 1.10 & $1.10$ \\
 $15\,716.252$ & $1.60(51)$ & $1.3960$  & 1.3802  & $1.7039$  & $1.15$ & 1.16 & $0.94$ \\
 $16\,506.066$ & $1.28(50)$ &           & 1.1422  & $1.2165$  &        & 1.12 & $1.05$ \\
 $16\,660.069$ & $0.38(19)$ &           & 0.4631  & $0.5872$  &        & 0.82 & $0.65$\\
\hline\hline
\multicolumn{8}{l}{\parbox{\linewidth}{\vspace{3pt}$\hphantom{I}^*${\footnotesize The assignment of this observed line is doubtful as its intensity is poorly predicted by all theoretical calculations; it was not included in the calculation of standard deviations.}}}
\end{tabular}}
\label{tab_DMS}
\end{center}
\end{table}
\normalsize

The MiZATeP line list has been compared directly with the NMT one. This
comparison shows good coincidence between the two; for example,
at room temperature the standard deviation of the ratio of Einstein's A coefficients 
of the 292 strongest lines (with relative intensity values greater than 0.001) 
from these line lists is only about 3\%.

\section{Partition function and intensity calculations}\label{part_func}
The labelling procedures described in the previous section
were used to assign statistical weights to
the line list transitions and for the calculation of the cooling and partition function.
In all these cases we used the second set of
nuclear motion calculations, which have comparatively low accuracy, to
supplement our high-accuracy levels with levels with energies
between 25\,000~cm$^{-1}$ and dissociation. This is essential to
obtain an accurate partition function at high temperatures.  We
used the same analytical form for the partition function as \citet{jt169}.

Our estimates show that the low accuracy energy levels in the second set of 
nuclear motion calculations as well as the absence of exact labelling procedure 
in this case influences the partition function values only slightly: the relative
error is less than $10^{-5}$ for each term in the partition function sum
and therefore we can safely ignore this effect. 

We computed a number of partition functions. In particular, $Q_{37}$
sums over the levels given in our final states file, which contains
levels with $J \leq 37$ and $E$ at least up to 35~000 \cm. $Q_{37}$ is therefore
consistent with the associated transitions file.  Other partition
sums, denoted $Q_J$, which summed levels up to $J$ and $E \leq$ 25~000
\cm\ were also computed.  Finally, a partition function computed by
summing over all levels for which we calculated energies is denoted
$Q_{\rm all}$.  $Q_{\rm all}$ provides a measure of convergence for
the other partition functions which sum over fewer levels.

Table~\ref{tab_part_func} gives our partition function results. It
compares our best estimates ($Q_{\rm all}$ and $Q_{37}$) with value by
\citet{jt169} and our more approximate sums. While the various values
agree well for lower temperatures, our most complete calculations give
significantly higher values at high $T$.  This suggests that the
partition function of H$_3^+$ has thus far been underestimated for temperatures
above 2000~K.

\begin{table}
\begin{center}
\caption{Partition function values, $Q$, as a function of temperature, $T$.
$Q_{\rm NT}$ are the values of \citet{jt169}; while $Q_J$ are our
values summed up to $J=20$ and 25\,000~cm$^{-1}$, 
$J=25$ and 25\,000~cm$^{-1}$, $J=37$ and 35\,000~cm$^{-1}$ (based on our states file); $Q_{\rm all}$ denotes 
partition function values obtained using all calculated energy states with $J$ 
up to $40$ and energies up to 42\,000~cm$^{-1}$.}
{\setlength{\tabcolsep}{3pt}
\begin{tabular}{rrrrrr}
\hline\hline
$T$ (K) & $Q_{\rm NT}$    & $Q_{20}$        & $Q_{25}$        & $Q_{37}$        & $Q_{\rm all}$       \\
\hline
 $100$  & $7.360$         & $7.397$         & $7.397$         & $7.397$         & $7.397$         \\
 $500$  & $80.579$        & $80.581$        & $80.581$        & $80.581$        & $80.581$        \\
 $1000$ & $245.762$       & $245.774$       & $245.775$       & $245.775$       & $245.775$       \\
 $1400$ & $473.731$       & $473.833$       & $473.875$       & $473.875$       & $473.875$       \\
 $2000$ & $1102.926$      & $1106.588$      & $1108.442$      & $1108.539$      & $1108.539$      \\
 $2400$ & $1808.406$      & $1832.712$      & $1842.438$      & $1843.513$      & $1843.514$      \\
 $3000$ & $3438.088$      & $3623.212$      & $3682.579$      & $3698.207$      & $3698.310$      \\ 
 $3500$ & $5385.317$      & $6005.538$      & $6186.521$      & $6268.304$      & $6269.639$      \\
 $4000$ & $7870.782$      & $9441.981$      & $9877.496$      & $10 \, 175.791$ & $10 \, 184.991$ \\
 $4500$ & $10 \, 851.290$ & $14 \, 134.011$ & $15 \, 018.507$ & $15 \, 857.630$ & $15 \, 899.213$ \\
 $5000$ & $14 \, 259.164$ & $20 \, 231.616$ & $21 \, 815.767$ & $23 \, 766.140$ & $23 \, 905.737$ \\
\hline\hline
\end{tabular}}
\label{tab_part_func}
\end{center}
\end{table}

The partition function $Q_{\rm all}$ provides our best estimate.  It
differs only slightly, the maximum difference is about 0.6\%\ at
5000~K, from $Q_{37}$ which was obtained using only our levels in our
final states file, as was our cooling function calculation. Energy
states with $J = 38 - 40$ are absent from the states file as they do
not participate in transitions with frequency values less than
25\,000~cm$^{-1}$.  The comparison of the partition functions suggests
that our line list and cooling function can be regarded as at least
99\%\ complete for temperatures up to 5000~K. 

We recommend using our partition function directly and note that
simply summing levels in the states file will give incorrect values
because of the duplicate
low-precision $J = 25$ levels present in this file. The partition
function and cooling function are given in steps of 1 K up to 5000 K
in the supplementary material.

\begin{figure}
\includegraphics[width=\columnwidth]{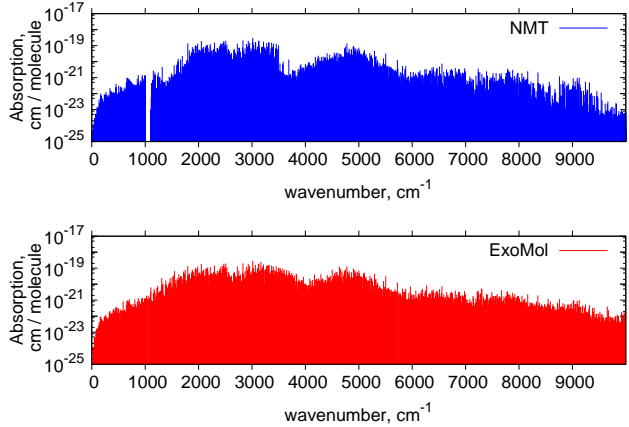}
\caption{Comparison of MiZATeP line list with the NMT one \citet{jt181} for the temperature value 2500~K.}
\label{NMT_comp_2500K}
\end{figure}

Figures \ref{NMT_comp_296K} and \ref{NMT_comp_2500K} compare the
MiZATeP and NMT line lists at room temperature and at 2500~K,
respectively, for the frequency range up to 10\,000~cm$^{-1}$. There is
generally good agreement although NMT appears to have an unexplained
gap in their data between 1000~cm$^{-1}$ to 1110~cm$^{-1}$ which is
not present in our new calculations. At room temperature the two line lists
give similar results, whereas at 2500~K there are obvious differences
between them.

\begin{figure}
\includegraphics[width=\columnwidth]{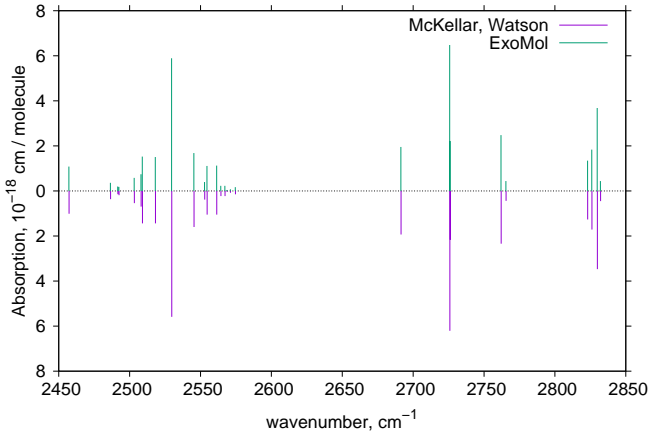}
\caption{Comparison of calculated spectral lines with the experimental ones obtained by \citet{98McWa}. 
The calculations were performed with temperature value equal to 285~K.}
\label{exp_comp_DMS7_285K}
\end{figure}

We compared the MiZATeP line list with the only available laboratory
measurement giving absolute transitions intensities, which was
performed by \citet{98McWa}.  To carry out this comparison it was necessary
to estimate the temperature of the observed spectrum; a value of 285~K
was chosen by inspection of the intensity ratios.  Figure
\ref{exp_comp_DMS7_285K} shows the result. The agreement is excellent,
with a standard deviation between the calculated intensity values from
experiment of about $6\%$; this difference probably reflects the
uncertainty in the assumed temperature and 
deviations from thermodynamic equilibrium in the experimental sample.

\begin{figure}
\includegraphics[width=\columnwidth]{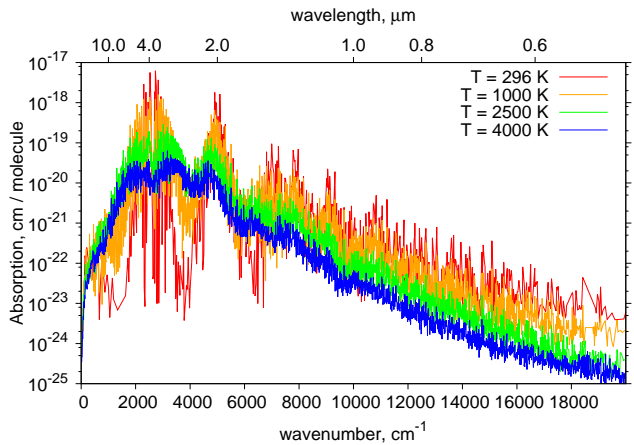}
\caption{Temperature dependence of MiZATeP line list for H$_3^+$. The curves become
increasingly smooth as the temperature increases.}
\label{T_depend}
\end{figure}

Finally, figure \ref{T_depend} illustrates temperature dependence of the MiZATeP line list over a wide temperature range: 
from room temperature to 4000~K. At the highest temperatures the absorption spectrum becomes much smoother.

\section{Lifetimes and cooling function calculations}\label{lifetime}
Lifetimes of states from the obtained list of energy levels
were computed. The algorithm of this calculation was standard
\citep{jt624}: we obtained a sum of Einstein $A$ coefficients of each
transition from our final line list, which includes the given level as
an upper one. The inverse value of the calculated sum is the
sought-for lifetime of the given state. Lifetimes were only
obtained for states for which accurate calculations were available:
those with $J$ up to 20 and energies less than 25\,000~cm$^{-1}$.

Our lifetimes calculations give an interesting result. Any molecular
system possesses a few very long-lived quantum states from which radiative decay
is impossible either because of the absence of lower-lying states, or
because such transitions are forbidden by selection rules. 
For example, a recent study on the H$_3$O$^+$ system
found 3 such metastable states for H$_3$O$^+$ and 4 for D$_3$O$^+$
\cite{jt660}.
We find a number of such states for which decay is not possible, all of which
belong to the vibrational ground state of the system.
Considering states with $J \leq 19$, we find a total 17 stable states
for the H$_3^+$ system, with energies up to 8509~cm$^{-1}$.  These
states are listed in table \ref{stab_states}. Only a few
(meta-)stable states could be anticipated on symmetry grounds.
The other states are stabilized because there are no lower-lying
states (generally levels in the $J-1$ manifold), which are reachable
given the rather stringent selection rules in force in H$_3^+$.
These metastable states are responsible for the observed astrophysical
and laboratory lifetime effects discussed in the Introduction.

The new line list was used to compute cooling function values for
temperatures up to 5000~K. The cooling function is the total energy
emitted by a single molecule in one second per unit solid angle.  We
used the analytical form given by \citet{jt624} and a version of
states file with purely calculated energies (i.e., without replacing them by
MARVEL analysis results) to compute the cooling function.

Table~\ref{tab_cool_func} gives our cooling function results. It
compares them ($W_{our}$) with values $W_{\rm Mel}$ from
\citet{Melin2006} and $W_{\rm Mil}$ from \citet{jt551} when possible
(the cooling curve presented in \citet{Melin2006} is valid only in
temperature range from 500 to 1800~K, while the one from \citet{jt551}
can be calculated for temperature values 30 -- 5000~K). The standard
deviation of the ratio of our results to the ones by \citet{Melin2006}
is about $33\%$, while for comparison with \citet{jt551} its value is
about $43\%$.

\begin{table}
\begin{center}
\caption{Cooling function values, $W$, as a function of temperature, $T$.
$W_{\rm Mel}$ and $W_{\rm Mil}$ are the values of \citet{Melin2006} and 
\citet{jt551}, respectively, in units of Watts Molecule$^{-1}$ Sterradian$^{-1}$, 
while $W_{our}$ are our values, in the same units system, 
summed up to $J=37$ and 42\,000~cm$^{-1}$ (based on our states file).}
\small
{\setlength{\tabcolsep}{3pt}
\begin{tabular}{rrrrrr}
\hline\hline
$T$ (K) & $W_{\rm Mel}$ & $W_{\rm Mil}$ & $W_{our}$ & \
$\frac{W_{our}}{W_{\rm Mel}}$ & $\frac{W_{our}}{W_{\rm Mil}}$ \\
\hline
  $20$  &                      &                      & $4.43\times10^{-32}$ &          &         \\
  $50$  &                      & $3.36\times10^{-30}$ & $3.37\times10^{-30}$ &          & $1.003$ \\
 $100$  &                      & $1.29\times10^{-28}$ & $1.26\times10^{-28}$ &          & $0.977$ \\
 $150$  &                      & $1.01\times10^{-27}$ & $1.03\times10^{-27}$ &          & $1.020$ \\
 $200$  &                      & $1.63\times10^{-26}$ & $1.69\times10^{-26}$ &          & $1.037$ \\
 $300$  &                      & $5.35\times10^{-24}$ & $5.32\times10^{-24}$ &          & $0.994$ \\
 $500$  & $5.05\times10^{-22}$ & $6.77\times10^{-22}$ & $6.69\times10^{-22}$ & $1.325$  & $0.988$ \\
 $700$  & $4.16\times10^{-21}$ & $5.74\times10^{-21}$ & $5.52\times10^{-21}$ & $1.327$  & $0.962$ \\
 $900$  & $1.41\times10^{-20}$ & $2.05\times10^{-20}$ & $1.87\times10^{-20}$ & $1.326$  & $0.912$ \\
 $1200$ & $4.49\times10^{-20}$ & $7.45\times10^{-20}$ & $5.95\times10^{-20}$ & $1.325$  & $0.799$ \\
 $1500$ & $9.80\times10^{-20}$ & $1.92\times10^{-19}$ & $1.30\times10^{-19}$ & $1.327$  & $0.677$ \\
 $1700$ & $1.47\times10^{-19}$ & $3.21\times10^{-19}$ & $1.95\times10^{-19}$ & $1.327$  & $0.607$ \\
 $1800$ & $1.75\times10^{-19}$ & $4.03\times10^{-19}$ & $2.33\times10^{-19}$ & $1.331$  & $0.578$ \\
 $2000$ &                      & $6.05\times10^{-19}$ & $3.20\times10^{-19}$ &          & $0.529$ \\
 $3000$ &                      & $2.16\times10^{-18}$ & $9.59\times10^{-19}$ &          & $0.444$ \\
 $4000$ &                      & $3.81\times10^{-18}$ & $1.80\times10^{-18}$ &          & $0.472$ \\
 $5000$ &                      & $4.77\times10^{-18}$ & $2.63\times10^{-18}$ &          & $0.551$ \\
\hline\hline
\end{tabular}}
\label{tab_cool_func}
\end{center}
\end{table}
\normalsize

\begin{table}
\caption{Calculated H$_3^+$ energy states with infinite lifetimes, $E_{\rm calc}$, together with corresponding energy levels, $E_{\rm M}$, obtained during the MARVEL analysis by \citet{13FuSzMa.H3+}.}
\small
{\setlength{\tabcolsep}{2pt}
\begin{tabular}{rrrrrrrrrrrr}
\hline\hline
$n$ & $E_{\rm calc}$ & $E_{\rm M}$ & $\Delta$ & sym & $\nu_1$ & $\nu_2$ & $l_2$ & $J$ & $G$ & $U$ & $K$ 
\\
\hline
1  & 64.12331   & 64.121000   & 50.0    & $E''$   & 0 & 0 & 0 & 1  & 1  & m & 1  \\
2  & 86.96619   & 86.960000   & 50.0    & $A_2'$  & 0 & 0 & 0 & 1  & 0  & m & 0  \\
3  & 315.31645  & 315.354081  & 15.2    & $A_2''$ & 0 & 0 & 0 & 3  & 3  & m & 3  \\
4  & 995.72428  & 995.890624  & 507.8   & $A_2'$  & 0 & 0 & 0 & 6  & 6  & m & 6  \\
5  & 1301.93329 & 1302.142000 & 10100.0 & $E''$   & 0 & 0 & 0 & 7  & 7  & m & 7  \\
6  & 2030.26910 & 2030.625886 & 833.3   & $A_2''$ & 0 & 0 & 0 & 9  & 9  & m & 9  \\
7  & 2451.10129 &             &         & $E'$    & 0 & 0 & 0 & 10 & 10 & m & 10 \\
8  & 2856.41347 & 2856.730003 & 1111.1  & $A_2''$ & 0 & 0 & 0 & 10 & 9  & m & 9  \\
9  & 3402.42821 &             &         & $A_2'$  & 0 & 0 & 0 & 12 & 12 & m & 12 \\
10 & 3931.31406 &             &         & $E''$   & 0 & 0 & 0 & 13 & 13 & m & 13 \\
11 & 4449.14478 &             &         & $A_2'$  & 0 & 0 & 0 & 13 & 12 & m & 12 \\
12 & 5091.29170 &             &         & $A_2''$ & 0 & 0 & 0 & 15 & 15 & m & 15 \\
13 & 5720.68071 &             &         & $E'$    & 0 & 0 & 0 & 16 & 16 & m & 16 \\
14 & 6341.32985 &             &         & $A_2''$ & 0 & 0 & 0 & 16 & 15 & m & 15 \\
15 & 7074.35983 &             &         & $A_2'$  & 0 & 0 & 0 & 18 & 18 & m & 18 \\
16 & 7797.41071 &             &         & $E''$   & 0 & 0 & 0 & 19 & 19 & m & 19 \\
17 & 8508.15437 &             &         & $A_2'$  & 0 & 0 & 0 & 19 & 18 & m & 18 \\
\hline\hline
\end{tabular}}
\label{stab_states}
\normalsize

\mbox{}\\
{\flushleft
$n$:   State counting number.     \\
$E_{\rm calc}$/$E_{\rm M}$: Calculated here/MARVEL state energy in cm$^{-1}$. \\
$\Delta$: Uncertainty of MARVEL energy states in 10$^{-6}$cm$^{-1}$.\\
sym: Symmetry of the state. \\
$\nu_1$: Symmetric stretch quantum number.\\  
$\nu_2$: Bending quantum number.\\ 
$l_2$: Vibrational angular momentum quantum number of the degenerate $\nu_2$ mode.\\
$J$: Total angular momentum.            \\
$K$: Absolute value of the projection of $J$ on the C$_3$.\\
$G$: Absolute value of quantum number $g = k - l_2$ \citep{84Watson.H3+}.\\
$U$: $U$-notation of \citet{84Watson.H3+}. \\ 
}

\end{table}

\begin{figure}
\includegraphics[width=\columnwidth]{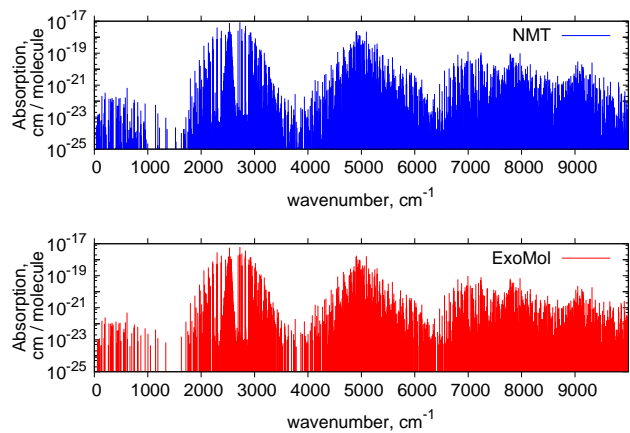}
\caption{Comparison of MiZATeP line list with the NMT one \citet{jt181} for the room temperature 296~K.}
\label{NMT_comp_296K}
\end{figure}

\begin{table}
\caption{ Extract from the states file for H$_3^+$. The full table is available 
from
http://cdsarc.u-strasbg.fr/cgi-bin/VizieR?-source=J/MNRAS/xxx/yy.}
\small
{\setlength{\tabcolsep}{2.5pt}
\begin{tabular}{rrrrrrrrrrrrr}
\hline\hline
$i$ & $\tilde{E}$& $g$ & $J$ & $\tau$ & $p$ & sym & $\nu_1$ & $\nu_2$ & $l_2$ & $G$ & $U$ & $K$ \\
\hline
           1  &    0.000000   &   0   &   0 &         NaN & e &  $A_1'$ & 0 & 0 & 0 & 0 & m & 0 \\
           2  &   64.121000   &   6   &   1 &         INF & e &  $E''$ & 0 & 0 & 0 & 1 & m & 1 \\
           3  &   86.960000   &  12   &   1 &         INF & f &  $A_2'$ & 0 & 0 & 0 & 0 & m & 0 \\
           4  &  169.294000   &  10   &   2 &  2.3491E+06 & e &  $E'$ & 0 & 0 & 0 & 2 & m & 2 \\
           5  &  237.357000   &  10   &   2 &  1.7812E+06 & f &  $E''$ & 0 & 0 & 0 & 1 & m & 1 \\
           6  &  315.354081   &  28   &   3 &         INF & e &  $A_2''$ & 0 & 0 & 0 & 3 & m & 3 \\
           7  &  428.019000   &  14   &   3 &  5.7399E+04 & f &  $E'$ & 0 & 0 & 0 & 2 & m & 2 \\
           8  &  494.773333   &  14   &   3 &  2.6579E+04 & e &  $E''$ & 0 & 0 & 0 & 1 & m & 1 \\ 
           9  &  502.028333   &  18   &   4 &  3.9059E+08 & e &  $E'$ & 0 & 0 & 0 & 4 & m & 4 \\
          10  &  516.878695   &  28   &   3 &  1.3589E+04 & f &  $A_2'$ & 0 & 0 & 0 & 0 & m & 0 \\
          11  &  658.722423   &  36   &   4 &  1.6935E+04 & f &  $A_2''$ & 0 & 0 & 0 & 3 & m & 3 \\
          12  &  729.031652   &  22   &   5 &  6.7686E+09 & e &  $E''$ & 0 & 0 & 0 & 5 & m & 5 \\
          13  &  768.475373   &  18   &   4 &  5.5360E+03 & e &  $E'$ & 0 & 0 & 0 & 2 & m & 2 \\
          14  &  833.578848   &  18   &   4 &  1.6480E+03 & f &  $E''$ & 0 & 0 & 0 & 1 & m & 1 \\
          15  &  928.965633   &  22   &   5 &  4.6803E+04 & f &  $E'$ & 0 & 0 & 0 & 4 & m & 4 \\
          16  &  995.890624   &  52   &   6 &         INF & e &  $A_2'$ & 0 & 0 & 0 & 6 & m & 6 \\
          17  & 1080.490719   &  44   &   5 &  5.5069E+04 & e &  $A_2''$ & 0 & 0 & 0 & 3 & m & 3 \\
          18  & 1187.117384   &  22   &   5 &  4.9087E+02 & f &  $E'$ & 0 & 0 & 0 & 2 & m & 2 \\
          19  & 1238.467378   &  26   &   6 &  1.5981E+05 & f &  $E''$ & 0 & 0 & 0 & 5 & m & 5 \\
          20  & 1250.313955   &  22   &   5 &  3.0108E+02 & e &  $E''$ & 0 & 0 & 0 & 1 & m & 1 \\
\hline\hline
\end{tabular}}
\label{tab:states}
\normalsize

\mbox{}\\
{\flushleft
$i$:   State counting number.     \\
$\tilde{E}$: State energy in \cm. \\
$g$: Total degeneracy of the state.\\
$J$: Total angular momentum.            \\
$\tau$: Lifetime of the state. INF means that the given state is metastable, NaN denotes unknown lifetime values of states without accurate labelling. \\  
$p$: e/f -- parity as given by DVR3D \cite{jt338}. \\
sym: Symmetry of the state. \\
$\nu_1$: Symmetric stretch quantum number.\\  
$\nu_2$: Bending quantum number.\\ 
$l_2$: Vibrational angular momentum quantum number of the degenerate $\nu_2$ mode.\\
$J$: Total angular momentum.            \\
$K$: Absolute value of the projection of $J$ on the C$_3$.\\
$G$: Absolute value of quantum number $g = k - l_2$ \citep{84Watson.H3+}.\\
$U$: $U$-notation of \citet{84Watson.H3+}.\\ 
}

\end{table}

\begin{table}
\caption{Extract from the transitions file for H$_3^+$.
The full table is available from
http://cdsarc.u-strasbg.fr/cgi-bin/VizieR?-source=J/MNRAS/xxx/yy.}
\begin{tabular}{rrr}
\hline\hline
$i$ & $f$ & $A_{if}$ \\
\hline
       55649   &     55648 & 1.7919E-16 \\
       42887   &     42882 & 2.2552E-13 \\
       85624   &     85623 & 4.3421E-25 \\
       88580   &     88579 & 1.5729E-22 \\
       55549   &     55548 & 3.6088E-13 \\
       46682   &     46681 & 4.3625E-14 \\
       62743   &     62742 & 3.4064E-14 \\
       55021   &     55017 & 5.8630E-14 \\
       59376   &     59371 & 4.7837E-13 \\
       31241   &     31239 & 1.5502E-12 \\
      100507   &    100506 & 9.0073E-22 \\
       28798   &     28795 & 3.3924E-12 \\
       82321   &     82320 & 1.6180E-20 \\
       81287   &     81282 & 2.0435E-12 \\
       68802   &     68801 & 1.9590E-13 \\
       98580   &     98579 & 3.8420E-20 \\
       70437   &     70436 & 8.0826E-24 \\
       47335   &     47334 & 2.8127E-13 \\
       80312   &     80308 & 6.5889E-15 \\
       60950   &     60949 & 6.0748E-20 \\
\hline\hline
\end{tabular}
\label{tab:trans}

\mbox{}\\
{\flushleft
$i$: Upper  state counting number. \\
$f$:  Lower  state counting number. \\
$A_{if}$:  Einstein-A coefficient in s$^{-1}$. \\
}
\end{table}

\section{Conclusion}
The MiZATeP full line list can be downloaded from the CDS, via
\url{ftp://cdsarc.u-strasbg.fr/pub/cats/J/MNRAS/xxx/yy}, or
\url{http://cdsarc.u-strasbg.fr/viz-bin/qcat?J/MNRAS//xxx/yy}, as well
as the exomol website, \url{www.exomol.com}. The line lists, cooling and
partition functions together with auxiliary data including the
potential parameters and dipole moment functions
can all be obtained also from \url{www.exomol.com} as part of the extended
ExoMol database \citep{jt631}.

\section*{Acknowledgements}
This work was supported by the ERC under the Advanced Investigator Project 267219 and 
by the Russian Fund for Fundamental Research (grant 15-02-07473).
\bibliographystyle{mnras}

\begin{thebibliography}{}
\makeatletter
\relax
\def\mn@urlcharsother{\let\do\@makeother \do\$\do\&\do\#\do\^\do\_\do\%\do\~}
\def\mn@doi{\begingroup\mn@urlcharsother \@ifnextchar [ {\mn@doi@}
  {\mn@doi@[]}}
\def\mn@doi@[#1]#2{\def\@tempa{#1}\ifx\@tempa\@empty \href
  {http://dx.doi.org/#2} {doi:#2}\else \href {http://dx.doi.org/#2} {#1}\fi
  \endgroup}
\def\mn@eprint#1#2{\mn@eprint@#1:#2::\@nil}
\def\mn@eprint@arXiv#1{\href {http://arxiv.org/abs/#1} {{\tt arXiv:#1}}}
\def\mn@eprint@dblp#1{\href {http://dblp.uni-trier.de/rec/bibtex/#1.xml}
  {dblp:#1}}
\def\mn@eprint@#1:#2:#3:#4\@nil{\def\@tempa {#1}\def\@tempb {#2}\def\@tempc
  {#3}\ifx \@tempc \@empty \let \@tempc \@tempb \let \@tempb \@tempa \fi \ifx
  \@tempb \@empty \def\@tempb {arXiv}\fi \@ifundefined
  {mn@eprint@\@tempb}{\@tempb:\@tempc}{\expandafter \expandafter \csname
  mn@eprint@\@tempb\endcsname \expandafter{\@tempc}}}

\bibitem[\protect\citeauthoryear{Al-Refaie, Yurchenko, Yachmenev  \&
  Tennyson}{Al-Refaie et~al.}{2015}]{jt597}
Al-Refaie A.~F.,  Yurchenko S.~N.,  Yachmenev A.,   Tennyson J.,  2015, \mn@doi
  [MNRAS] {10.1093/mnras/stv091}, 448, 1704

\bibitem[\protect\citeauthoryear{Al-Refaie, Polyansky, I., Ovsyannikov,
  Tennyson  \& Yurchenko}{Al-Refaie et~al.}{2016}]{jt638}
Al-Refaie A.~F.,  Polyansky O.~L.,  I. R.,  Ovsyannikov Tennyson J.,
  Yurchenko S.~N.,  2016, \mn@doi [MNRAS] {10.1093/mnras/stw1295}, 461, 1012

\bibitem[\protect\citeauthoryear{Asvany, Hugo, Schlemmer, Muller, Kuhnemann,
  Schiller  \& Tennyson}{Asvany et~al.}{2007}]{jt413}
Asvany O.,  Hugo E.,  Schlemmer S.,  Muller F.,  Kuhnemann F.,  Schiller S.,
  Tennyson J.,  2007, J. Chem. Phys., 127, 154317

\bibitem[\protect\citeauthoryear{Azzam, Yurchenko, Tennyson  \& Naumenko}{Azzam
  et~al.}{2016}]{jt640}
Azzam A. A.~A.,  Yurchenko S.~N.,  Tennyson J.,   Naumenko O.~V.,  2016,
  \mn@doi [MNRAS] {10.1093/mnras/stw1133}, 460, 4063

\bibitem[\protect\citeauthoryear{Barber, Tennyson, Harris  \& Tolchenov}{Barber
  et~al.}{2006}]{jt378}
Barber R.~J.,  Tennyson J.,  Harris G.~J.,   Tolchenov R.~N.,  2006, MNRAS,
  368, 1087

\bibitem[\protect\citeauthoryear{Barber, Strange, Hill, Polyansky, Mellau,
  Yurchenko  \& Tennyson}{Barber et~al.}{2014}]{jt570}
Barber R.~J.,  Strange J.~K.,  Hill C.,  Polyansky O.~L.,  Mellau G.~C.,
  Yurchenko S.~N.,   Tennyson J.,  2014, \mn@doi [MNRAS]
  {10.1093/mnras/stt2011}, 437, 1828

\bibitem[\protect\citeauthoryear{Barton, Yurchenko  \& Tennyson}{Barton
  et~al.}{2013}]{jt563}
Barton E.~J.,  Yurchenko S.~N.,   Tennyson J.,  2013, MNRAS, 434, 1469

\bibitem[\protect\citeauthoryear{Barton, Chiu, Golpayegani, Yurchenko,
  Tennyson, Frohman  \& Bernath}{Barton et~al.}{2014}]{jt583}
Barton E.~J.,  Chiu C.,  Golpayegani S.,  Yurchenko S.~N.,  Tennyson J.,
  Frohman D.~J.,   Bernath P.~F.,  2014, \mn@doi [MNRAS]
  {10.1093/mnras/stu944}, 442, 1821

\bibitem[\protect\citeauthoryear{Bergeron, Ruiz  \& Leggett}{Bergeron
  et~al.}{1997}]{brl97}
Bergeron P.,  Ruiz M.~T.,   Leggett S.~K.,  1997, ApJS, 108, 339

\bibitem[\protect\citeauthoryear{Brittain \& Rettig}{Brittain \&
  Rettig}{2002}]{02BrRexx.H3+}
Brittain S.~D.,  Rettig T.,  2002, \mn@doi [Nature] {{10.1038/nature00800}},
  418, 57

\bibitem[\protect\citeauthoryear{Cencek, Rychlewski, Jaquet  \&
  Kutzelnigg}{Cencek et~al.}{1998}]{Cencek1998}
Cencek W.,  Rychlewski J.,  Jaquet R.,   Kutzelnigg W.,  1998, J. Chem. Phys.,
  108, 2831

\bibitem[\protect\citeauthoryear{Coppola, Lodi  \& Tennyson}{Coppola
  et~al.}{2011}]{jt506}
Coppola C.~M.,  Lodi L.,   Tennyson J.,  2011, MNRAS, 415, 487

\bibitem[\protect\citeauthoryear{Dinelli, Polyansky  \& Tennyson}{Dinelli
  et~al.}{1995}]{jt175}
Dinelli B.~M.,  Polyansky O.~L.,   Tennyson J.,  1995, J. Chem. Phys., 103,
  10433

\bibitem[\protect\citeauthoryear{Dinelli, Neale, Polyansky  \&
  Tennyson}{Dinelli et~al.}{1997}]{jt193}
Dinelli B.~M.,  Neale L.,  Polyansky O.~L.,   Tennyson J.,  1997, J. Mol.
  Spectrosc., 181, 142

\bibitem[\protect\citeauthoryear{Diniz, Mohallem, Alijah, Pavanello, Adamowicz,
  Polyansky  \& Tennyson}{Diniz et~al.}{2013}]{jt566}
Diniz L.~G.,  Mohallem J.~R.,  Alijah A.,  Pavanello M.,  Adamowicz L.,
  Polyansky O.~L.,   Tennyson J.,  2013, \mn@doi [Phys. Rev. A]
  {10.1103/PhysRevA.88.032506}, 88, 032506

\bibitem[\protect\citeauthoryear{Drossart et~al.,}{Drossart
  et~al.}{1989}]{jt80}
Drossart P.,  et~al., 1989, Nature, 340, 539

\bibitem[\protect\citeauthoryear{Engel, Doss, Harris  \& Tennyson}{Engel
  et~al.}{2005}]{jt347}
Engel E.~A.,  Doss N.,  Harris G.~J.,   Tennyson J.,  2005, MNRAS, 357, 471

\bibitem[\protect\citeauthoryear{Farnik, Davis, Kostin, Polyansky, Tennyson  \&
  Nesbitt}{Farnik et~al.}{2002}]{jt289}
Farnik M.,  Davis S.,  Kostin M.~A.,  Polyansky O.~L.,  Tennyson J.,   Nesbitt
  D.~J.,  2002, J. Chem. Phys., 116, 6146

\bibitem[\protect\citeauthoryear{Furtenbacher \& {Cs\'asz\'ar}}{Furtenbacher \&
  {Cs\'asz\'ar}}{2012}]{12FuCs.method}
Furtenbacher T.,  {Cs\'asz\'ar} A.~G.,  2012, J. Quant. Spectrosc. Radiat.
  Transf., 113, 929

\bibitem[\protect\citeauthoryear{Furtenbacher, {Cs\'asz\'ar}  \&
  Tennyson}{Furtenbacher et~al.}{2007}]{jt412}
Furtenbacher T.,  {Cs\'asz\'ar} A.~G.,   Tennyson J.,  2007, J. Mol.
  Spectrosc., 245, 115

\bibitem[\protect\citeauthoryear{Furtenbacher, Szidarovszky, Matyus, Fabri  \&
  Csaszar}{Furtenbacher et~al.}{2013}]{13FuSzMa.H3+}
Furtenbacher T.,  Szidarovszky T.,  Matyus E.,  Fabri C.,   Csaszar A.~G.,
  2013, \mn@doi [J. Chem. Theor. Comput.] {10.1021/ct4004355}, 9, 5471

\bibitem[\protect\citeauthoryear{Geballe \& Oka}{Geballe \& Oka}{1996}]{g96}
Geballe T.~R.,  Oka T.,  1996, Nature, 384, 334

\bibitem[\protect\citeauthoryear{Geballe, Jagod  \& Oka}{Geballe
  et~al.}{1993}]{gj93}
Geballe T.~R.,  Jagod M.~F.,   Oka T.,  1993, ApJ, 408, L109

\bibitem[\protect\citeauthoryear{Geballe, Goto, Usuda, Oka  \& McCall}{Geballe
  et~al.}{2006}]{06GeGoUs.H3+}
Geballe T.~R.,  Goto M.,  Usuda T.,  Oka T.,   McCall B.~J.,  2006, ApJ, 644,
  907

\bibitem[\protect\citeauthoryear{Geballe, Mason  \& Oka}{Geballe
  et~al.}{2015}]{15GeMaOk.H3+}
Geballe T.~R.,  Mason R.~E.,   Oka T.,  2015, ApJ, 812, 56

\bibitem[\protect\citeauthoryear{Glover \& Savin}{Glover \&
  Savin}{2006}]{06GlSa}
Glover S.,  Savin D.~W.,  2006, \mn@doi [Phil. Trans. R. Soc. A]
  {10.1098/rsta.2006.1867}, 364, 3107

\bibitem[\protect\citeauthoryear{Goto, McCall, Geballe, Usuda, Kobayashi,
  Terada  \& Oka}{Goto et~al.}{2002}]{02GoMcGe.H3+}
Goto M.,  McCall B.~J.,  Geballe T.~R.,  Usuda T.,  Kobayashi N.,  Terada H.,
  Oka T.,  2002, PASJ, 54, 951

\bibitem[\protect\citeauthoryear{Goto, Geballe, McCall, Usuda, Suto, Terada,
  Kobayashi  \& Oka}{Goto et~al.}{2005}]{05GoGeMc.H3+}
Goto M.,  Geballe T.~R.,  McCall B.~J.,  Usuda T.,  Suto H.,  Terada H.,
  Kobayashi N.,   Oka T.,  2005, ApJ, 629, 865

\bibitem[\protect\citeauthoryear{Goto et~al.,}{Goto
  et~al.}{2008}]{08GoUsNa.H3+}
Goto M.,  et~al., 2008, \mn@doi [ApJ] {10.1086/591657}, 688, 306

\bibitem[\protect\citeauthoryear{Harris, Larner, Tennyson, Kaminsky, Pavlenko
  \& Jones}{Harris et~al.}{2008}]{jt447}
Harris G.~J.,  Larner F.~C.,  Tennyson J.,  Kaminsky B.~M.,  Pavlenko Y.~V.,
  Jones H. R.~A.,  2008, MNRAS, 390, 143

\bibitem[\protect\citeauthoryear{Herbst \& Klemperer}{Herbst \&
  Klemperer}{1973}]{hk73}
Herbst E.,  Klemperer W.,  1973, ApJ, 185, 505

\bibitem[\protect\citeauthoryear{Indriolo \& McCall}{Indriolo \&
  McCall}{2012}]{12InMc}
Indriolo N.,  McCall B.~J.,  2012, \mn@doi [ApJ] {10.1088/0004-637X/745/1/91},
  745, 91

\bibitem[\protect\citeauthoryear{Jaquet \& {Carrington}}{Jaquet \&
  {Carrington}}{2013}]{13JaCa}
Jaquet R.,  {Carrington} T.,  2013, \mn@doi [J. Phys. Chem. A]
  {10.1021/jp312027s}, 117, 9493

\bibitem[\protect\citeauthoryear{Johnson}{Johnson}{1983}]{JOH83:1916}
Johnson B.~R.,  1983, J. Chem. Phys., 79, 1916

\bibitem[\protect\citeauthoryear{Kao, Oka, Miller  \& Tennyson}{Kao
  et~al.}{1991}]{jt107}
Kao L.,  Oka T.,  Miller S.,   Tennyson J.,  1991, ApJS, 77, 317

\bibitem[\protect\citeauthoryear{Khodachenko, Shaikhislamov, Lammer  \&
  Prokopov}{Khodachenko et~al.}{2015}]{15KhShLa.H3+}
Khodachenko M.~L.,  Shaikhislamov I.~F.,  Lammer H.,   Prokopov P.~A.,  2015,
  \mn@doi [ApJ] {10.1088/0004-637X/813/1/50}, 813, 50

\bibitem[\protect\citeauthoryear{Koskinen, Aylward  \& Miller}{Koskinen
  et~al.}{2007}]{kam07}
Koskinen T.~T.,  Aylward A.~D.,   Miller S.,  2007, Nature, 450, 845

\bibitem[\protect\citeauthoryear{Kreckel et~al.,}{Kreckel et~al.}{2002}]{jt306}
Kreckel H.,  et~al., 2002, Phys. Rev. A, 66, 052509

\bibitem[\protect\citeauthoryear{Kreckel, Schwalm, Tennyson, Wolf  \&
  Zajfman}{Kreckel et~al.}{2004}]{jt340}
Kreckel H.,  Schwalm D.,  Tennyson J.,  Wolf A.,   Zajfman D.,  2004, New J.
  Phys, 6, 151

\bibitem[\protect\citeauthoryear{Kutzelnigg \& Jaquet}{Kutzelnigg \&
  Jaquet}{2006}]{kutzelnigg:2006}
Kutzelnigg W.,  Jaquet R.,  2006, Phil. Trans. R. Soc. A, 364, 2855

\bibitem[\protect\citeauthoryear{Lam, Achilleos, Miller, Tennyson, Trafton,
  Geballe  \& Ballester}{Lam et~al.}{1997a}]{jt201}
Lam H.~A.,  Achilleos N.,  Miller S.,  Tennyson J.,  Trafton L.~M.,  Geballe
  T.~R.,   Ballester G.~E.,  1997a, Icarus, 127, 379

\bibitem[\protect\citeauthoryear{Lam, Miller, Joseph, Geballe, Trafton,
  Tennyson  \& Ballester}{Lam et~al.}{1997b}]{jt192}
Lam H.~A.,  Miller S.,  Joseph R.~D.,  Geballe T.~R.,  Trafton L.~M.,  Tennyson
  J.,   Ballester G.~E.,  1997b, ApJ, 474, L73

\bibitem[\protect\citeauthoryear{Lie \& Frye}{Lie \& Frye}{1992}]{lf92}
Lie G.~C.,  Frye D.,  1992, J. Chem. Phys., 96, 6784

\bibitem[\protect\citeauthoryear{Lindsay \& McCall}{Lindsay \&
  McCall}{2001}]{lindsay:2001a}
Lindsay C.~M.,  McCall B.~J.,  2001, J. Mol. Spectrosc., 210, 60

\bibitem[\protect\citeauthoryear{McCall \& Oka}{McCall \& Oka}{2000}]{MO00}
McCall B.~J.,  Oka T.,  2000, Science, 287, 1941

\bibitem[\protect\citeauthoryear{McCall, Geballe, Hinkle  \& Oka}{McCall
  et~al.}{1998}]{mg98}
McCall B.~J.,  Geballe T.~R.,  Hinkle K.~H.,   Oka T.,  1998, Science, 279,
  1910

\bibitem[\protect\citeauthoryear{McCall, Geballe, Hinkle  \& Oka}{McCall
  et~al.}{1999}]{mccall:1999}
McCall B.~J.,  Geballe T.~R.,  Hinkle K.~H.,   Oka T.,  1999, ApJ, 522, 338

\bibitem[\protect\citeauthoryear{McCall et~al.,}{McCall et~al.}{2002}]{mh02}
McCall B.~J.,  et~al., 2002, Astrophys. J., 567, 391

\bibitem[\protect\citeauthoryear{McCall et~al.,}{McCall
  et~al.}{2003}]{mccall:2003b}
McCall B.~J.,  et~al., 2003, Nature, 422, 500

\bibitem[\protect\citeauthoryear{McKellar \& Watson}{McKellar \&
  Watson}{1998}]{98McWa}
McKellar A. R.~W.,  Watson J. K.~G.,  1998, J. Mol. Spectrosc., 191, 215

\bibitem[\protect\citeauthoryear{McKemmish, Yurchenko  \& Tennyson}{McKemmish
  et~al.}{2016}]{jt644}
McKemmish L.~K.,  Yurchenko S.~N.,   Tennyson J.,  2016, \mn@doi [MNRAS]
  {10.1093/mnras/stw1969}, 463, 771

\bibitem[\protect\citeauthoryear{Melin}{Melin}{2006}]{Melin2006}
Melin H.,  2006, PhD thesis, University College London

\bibitem[\protect\citeauthoryear{Melnikov, Yurchenko, Tennyson  \&
  Jensen}{Melnikov et~al.}{2016}]{jt660}
Melnikov V.~V.,  Yurchenko S.~N.,  Tennyson J.,   Jensen P.,  2016, \mn@doi
  [Phys. Chem. Chem. Phys.] {10.1039/C6CP04661D}, 18, 26268

\bibitem[\protect\citeauthoryear{Millar}{Millar}{2015}]{15Millar}
Millar T.~J.,  2015, \mn@doi [Plasma Sources Sci. Technol.]
  {{10.1088/0963-0252/24/4/043001}}, 24, 043001

\bibitem[\protect\citeauthoryear{Miller \& Tennyson}{Miller \&
  Tennyson}{1988a}]{jt64}
Miller S.,  Tennyson J.,  1988a, Chem. Phys. Lett., 145, 117

\bibitem[\protect\citeauthoryear{Miller \& Tennyson}{Miller \&
  Tennyson}{1988b}]{jt72}
Miller S.,  Tennyson J.,  1988b, ApJ, 335, 486

\bibitem[\protect\citeauthoryear{Miller, Tennyson, Lepp  \& Dalgarno}{Miller
  et~al.}{1992}]{jt110}
Miller S.,  Tennyson J.,  Lepp S.,   Dalgarno A.,  1992, Nature, 355, 420

\bibitem[\protect\citeauthoryear{Miller, Lam  \& Tennyson}{Miller
  et~al.}{1994}]{jt155}
Miller S.,  Lam H.~A.,   Tennyson J.,  1994, Can. J. Phys., 72, 760

\bibitem[\protect\citeauthoryear{Miller et~al.,}{Miller et~al.}{1995}]{jt167}
Miller S.,  et~al., 1995, Geophys. Res. Lett., 22, 1629

\bibitem[\protect\citeauthoryear{Miller et~al.,}{Miller et~al.}{2000}]{jt258}
Miller S.,  et~al., 2000, Phil. Trans. Royal Soc. London A, 358, 2485

\bibitem[\protect\citeauthoryear{Miller, Stallard, Melin  \& Tennyson}{Miller
  et~al.}{2010}]{jt489}
Miller S.,  Stallard T.,  Melin H.,   Tennyson J.,  2010, Faraday Discuss.,
  147, 283

\bibitem[\protect\citeauthoryear{Miller, Stallard, Tennyson  \& Melin}{Miller
  et~al.}{2013}]{jt551}
Miller S.,  Stallard T.,  Tennyson J.,   Melin H.,  2013, J. Phys. Chem. A,
  117, 9633

\bibitem[\protect\citeauthoryear{Moss}{Moss}{1996}]{96Moss}
Moss R.~E.,  {1996}, \mn@doi [Mol. Phys.] {{10.1080/002689796174083}}, {89},
  195

\bibitem[\protect\citeauthoryear{Munro, Ramanlal  \& Tennyson}{Munro
  et~al.}{2005}]{jt358}
Munro J.~J.,  Ramanlal J.,   Tennyson J.,  2005, New J. Phys, 7, 196

\bibitem[\protect\citeauthoryear{Neale \& Tennyson}{Neale \&
  Tennyson}{1995}]{jt169}
Neale L.,  Tennyson J.,  1995, ApJ, 454, L169

\bibitem[\protect\citeauthoryear{Neale, Miller  \& Tennyson}{Neale
  et~al.}{1996}]{jt181}
Neale L.,  Miller S.,   Tennyson J.,  1996, ApJ, 464, 516

\bibitem[\protect\citeauthoryear{Oka}{Oka}{2006}]{oka:2006}
Oka T.,  2006, PNAS, 103, 12235

\bibitem[\protect\citeauthoryear{Oka}{Oka}{2013}]{13Oka.H3+}
Oka T.,  2013, \mn@doi [Chem. Rev.] {10.1021/cr400266w}, 113, 8738

\bibitem[\protect\citeauthoryear{Oka, Geballe, Goto, Usuda  \& McCall}{Oka
  et~al.}{2005}]{05OkGeGo.H3+}
Oka T.,  Geballe T.~R.,  Goto M.,  Usuda T.,   McCall B.~J.,  2005, \mn@doi
  [ApJ] {10.1086/432679}, 632, 882

\bibitem[\protect\citeauthoryear{Pan \& Oka}{Pan \& Oka}{1986}]{86PaOk.H3+}
Pan F.~S.,  Oka T.,  1986, \mn@doi [ApJ] {10.1086/164264}, 305, 518

\bibitem[\protect\citeauthoryear{Patrascu, Tennyson  \& Yurchenko}{Patrascu
  et~al.}{2015}]{jt598}
Patrascu A.~T.,  Tennyson J.,   Yurchenko S.~N.,  2015, \mn@doi [MNRAS]
  {10.1093/mnras/stv507}, 449, 3613

\bibitem[\protect\citeauthoryear{Paulose, Barton, Yurchenko  \&
  Tennyson}{Paulose et~al.}{2015}]{jt615}
Paulose G.,  Barton E.~J.,  Yurchenko S.~N.,   Tennyson J.,  2015, \mn@doi
  [MNRAS] {10.1093/mnras/stv1543}, 454, 1931

\bibitem[\protect\citeauthoryear{Pavanello et~al.,}{Pavanello
  et~al.}{2012a}]{jt512}
Pavanello M.,  et~al., 2012a, Phys. Rev. Lett., 108, 023002

\bibitem[\protect\citeauthoryear{Pavanello et~al.,}{Pavanello
  et~al.}{2012b}]{jt526}
Pavanello M.,  et~al., 2012b, J. Chem. Phys., 136, 184303

\bibitem[\protect\citeauthoryear{Pavanello, Tung, Leonarski  \&
  Adamowicz}{Pavanello et~al.}{2009}]{ptf09}
Pavanello M.,  Tung W.-C.,  Leonarski F.,   Adamowicz L.,  {2009}, J. Chem.
  Phys., {130}, 074105

\bibitem[\protect\citeauthoryear{Pavlyuchko, Yurchenko  \& Tennyson}{Pavlyuchko
  et~al.}{2015}]{jt614}
Pavlyuchko A.~I.,  Yurchenko S.~N.,   Tennyson J.,  2015, \mn@doi [MNRAS]
  {10.1093/mnras/stv1376}, 452, 1702

\bibitem[\protect\citeauthoryear{Petrignani et~al.,}{Petrignani
  et~al.}{2014}]{jt587}
Petrignani A.,  et~al., 2014, \mn@doi [J. Chem. Phys.] {10.1063/1.4904440},
  141, 241104

\bibitem[\protect\citeauthoryear{Polyansky \& Tennyson}{Polyansky \&
  Tennyson}{1999}]{jt236}
Polyansky O.~L.,  Tennyson J.,  1999, J. Chem. Phys., 110, 5056

\bibitem[\protect\citeauthoryear{Polyansky, Dinelli, {Le Sueur}  \&
  Tennyson}{Polyansky et~al.}{1995}]{jt166}
Polyansky O.~L.,  Dinelli B.~M.,  {Le Sueur} C.~R.,   Tennyson J.,  1995, J.
  Chem. Phys., 102, 9322

\bibitem[\protect\citeauthoryear{Polyansky, Kyuberis, Lodi, Tennyson,
  Ovsyannikov  \& Zobov}{Polyansky et~al.}{2016}]{jt665}
Polyansky O.~L.,  Kyuberis A.~A.,  Lodi L.,  Tennyson J.,  Ovsyannikov R.~I.,
  Zobov N.,  2016, \mn@doi [MNRAS] {10.1093/mnras/stw3125}, 466, 1363

\bibitem[\protect\citeauthoryear{Polyansky, Kyuberis, Lodi, Tennyson,
  Ovsyannikov, Zobov  \& Yurchenko}{Polyansky et~al.}{2017}]{jtpoz}
Polyansky O.~L.,  Kyuberis A.~A.,  Lodi L.,  Tennyson J.,  Ovsyannikov R.~I.,
  Zobov N.,   Yurchenko S.~N.,  2017, MNRAS

\bibitem[\protect\citeauthoryear{Rego, Achilleos, Stallard, Miller, Prange,
  Dougherty  \& Joseph}{Rego et~al.}{1999}]{99ReAcSt.H3+}
Rego D.,  Achilleos N.,  Stallard T.,  Miller S.,  Prange R.,  Dougherty M.,
  Joseph R.~D.,  {1999}, Nature, {399}, 21

\bibitem[\protect\citeauthoryear{Rivlin, Lodi, Yurchenko, Tennyson  \& {Le
  Roy}}{Rivlin et~al.}{2015}]{jt605}
Rivlin T.,  Lodi L.,  Yurchenko S.~N.,  Tennyson J.,   {Le Roy} R.~J.,  2015,
  \mn@doi [MNRAS] {10.1093/mnras/stv979}, 451, 5153

\bibitem[\protect\citeauthoryear{R\"{o}hse, Kutzelnigg, Jaquet  \&
  Klopper}{R\"{o}hse et~al.}{1994}]{Rohse1994}
R\"{o}hse R.,  Kutzelnigg W.,  Jaquet R.,   Klopper W.,  1994, J. Chem. Phys.,
  101, 2231

\bibitem[\protect\citeauthoryear{Schiffels, Alijah  \& Hinze}{Schiffels
  et~al.}{2003b}]{SCH03:175}
Schiffels P.,  Alijah A.,   Hinze J.,  2003b, Mol. Phys., 101, 175

\bibitem[\protect\citeauthoryear{Schiffels, Alijah  \& Hinze}{Schiffels
  et~al.}{2003a}]{03ScAlHia.H3+}
Schiffels P.,  Alijah A.,   Hinze J.,  2003a, Mol. Phys., 101, 175

\bibitem[\protect\citeauthoryear{Schiffels, Alijah  \& Hinze}{Schiffels
  et~al.}{2003c}]{03ScAlHib.H3+}
Schiffels P.,  Alijah A.,   Hinze J.,  2003c, Mol. Phys., 101, 189

\bibitem[\protect\citeauthoryear{Shkolnik, Gaidos  \& Moskovitz}{Shkolnik
  et~al.}{2006}]{06ShGaMo.H3+}
Shkolnik E.,  Gaidos E.,   Moskovitz N.,  2006, \mn@doi [ApJ]
  {{10.1086/506476}}, 132, 1267

\bibitem[\protect\citeauthoryear{Sochi \& Tennyson}{Sochi \&
  Tennyson}{2010}]{jt478}
Sochi T.,  Tennyson J.,  2010, MNRAS, 405, 2345

\bibitem[\protect\citeauthoryear{Sousa-Silva, Hesketh, Yurchenko, Hill  \&
  Tennyson}{Sousa-Silva et~al.}{2014}]{jt571}
Sousa-Silva C.,  Hesketh N.,  Yurchenko S.~N.,  Hill C.,   Tennyson J.,  2014,
  \mn@doi [J. Quant. Spectrosc. Radiat. Transf.] {10.1016/j.jqsrt2014.03.012},
  142, 66

\bibitem[\protect\citeauthoryear{Sousa-Silva, Al-Refaie, Tennyson  \&
  Yurchenko}{Sousa-Silva et~al.}{2015}]{jt592}
Sousa-Silva C.,  Al-Refaie A.~F.,  Tennyson J.,   Yurchenko S.~N.,  2015,
  \mn@doi [MNRAS] {10.1093/mnras/stu2246}, 446, 2337

\bibitem[\protect\citeauthoryear{Stallard, Miller, Melin, Lystrup, Cowley,
  Bunce, Achilleos  \& Dougherty}{Stallard et~al.}{2008a}]{08StMiMe.H3+}
Stallard T.,  Miller S.,  Melin H.,  Lystrup M.,  Cowley S. W.~H.,  Bunce
  E.~J.,  Achilleos N.,   Dougherty M.,  {2008}a, \mn@doi [Nature]
  {{10.1038/nature07077}}, {453}, 1083

\bibitem[\protect\citeauthoryear{Stallard et~al.,}{Stallard
  et~al.}{2008b}]{08StMiLy.H3+}
Stallard T.,  et~al., {2008}b, \mn@doi [Nature] {{10.1038/nature07440}}, {456},
  214

\bibitem[\protect\citeauthoryear{Tennyson}{Tennyson}{1993}]{jt133}
Tennyson J.,  1993, J. Chem. Phys., 98, 9658

\bibitem[\protect\citeauthoryear{Tennyson}{Tennyson}{1995}]{jt157}
Tennyson J.,  1995, Rep. Prog. Phys., 58, 421

\bibitem[\protect\citeauthoryear{Tennyson \& Sutcliffe}{Tennyson \&
  Sutcliffe}{1982}]{jt14}
Tennyson J.,  Sutcliffe B.~T.,  1982, \mn@doi [J. Chem. Phys.]
  {10.1063/1.444316}, 77, 4061

\bibitem[\protect\citeauthoryear{Tennyson \& Sutcliffe}{Tennyson \&
  Sutcliffe}{1983}]{jt23}
Tennyson J.,  Sutcliffe B.~T.,  1983, J. Mol. Spectrosc., 101, 71

\bibitem[\protect\citeauthoryear{Tennyson \& Yurchenko}{Tennyson \&
  Yurchenko}{2012}]{jt528}
Tennyson J.,  Yurchenko S.~N.,  2012, \mn@doi [MNRAS]
  {10.1111/j.1365-2966.2012.21440.x}, 425, 21

\bibitem[\protect\citeauthoryear{Tennyson, Kostin, Barletta, Harris, Polyansky,
  Ramanlal  \& Zobov}{Tennyson et~al.}{2004}]{jt338}
Tennyson J.,  Kostin M.~A.,  Barletta P.,  Harris G.~J.,  Polyansky O.~L.,
  Ramanlal J.,   Zobov N.~F.,  2004, Comput. Phys. Commun., 163, 85

\bibitem[\protect\citeauthoryear{Tennyson, Hulme, Naim  \& Yurchenko}{Tennyson
  et~al.}{2016a}]{jt624}
Tennyson J.,  Hulme K.,  Naim O.~K.,   Yurchenko S.~N.,  2016a, \mn@doi [J.
  Phys. B: At. Mol. Opt. Phys.] {10.1088/0953-4075/49/4/044002}, 49, 044002

\bibitem[\protect\citeauthoryear{Tennyson et~al.,}{Tennyson
  et~al.}{2016b}]{jt631}
Tennyson J.,  et~al., 2016b, \mn@doi [J. Mol. Spectrosc.]
  {10.1016/j.jms.2016.05.002}, 327, 73

\bibitem[\protect\citeauthoryear{Trafton, Geballe, Miller, Tennyson  \&
  Ballester}{Trafton et~al.}{1993}]{jt127}
Trafton L.~M.,  Geballe T.~R.,  Miller S.,  Tennyson J.,   Ballester G.~E.,
  1993, ApJ, 405, 761

\bibitem[\protect\citeauthoryear{Underwood, Tennyson, Yurchenko, Huang,
  Schwenke, Lee, Clausen  \& Fateev}{Underwood et~al.}{2016a}]{jt635}
Underwood D.~S.,  Tennyson J.,  Yurchenko S.~N.,  Huang X.,  Schwenke D.~W.,
  Lee T.~J.,  Clausen S.,   Fateev A.,  2016a, \mn@doi [MNRAS]
  {10.1093/mnras/stw849}, 459, 3890

\bibitem[\protect\citeauthoryear{Underwood, Tennyson, Yurchenko, Clausen  \&
  Fateev}{Underwood et~al.}{2016b}]{jt641}
Underwood D.~S.,  Tennyson J.,  Yurchenko S.~N.,  Clausen S.,   Fateev A.,
  2016b, \mn@doi [MNRAS] {10.1093/mnras/stw1828}, 462, 4300

\bibitem[\protect\citeauthoryear{Voronin, Tennyson, Tolchenov, Lugovskoy  \&
  Yurchenko}{Voronin et~al.}{2010}]{jt469}
Voronin B.~A.,  Tennyson J.,  Tolchenov R.~N.,  Lugovskoy A.~A.,   Yurchenko
  S.~N.,  2010, MNRAS, 402, 492

\bibitem[\protect\citeauthoryear{Watson}{Watson}{1973}]{w73}
Watson W.~D.,  1973, ApJ, 183, L17

\bibitem[\protect\citeauthoryear{Watson}{Watson}{1984}]{84Watson.H3+}
Watson J. K.~G.,  {1984}, J. Mol. Spectrosc., {103}, 350

\bibitem[\protect\citeauthoryear{Whitten \& Smith}{Whitten \&
  Smith}{1968}]{WHI68:1103}
Whitten R.~C.,  Smith F.~T.,  1968, J. Math. Phys., 9, 1103

\bibitem[\protect\citeauthoryear{Wolniewicz}{Wolniewicz}{1988}]{WOL88:371}
Wolniewicz L.,  1988, J. Chem. Phys., 90, 371

\bibitem[\protect\citeauthoryear{Wolniewicz, Hinze  \& Alijah}{Wolniewicz
  et~al.}{1993}]{WOL93:2695}
Wolniewicz L.,  Hinze J.,   Alijah A.,  1993, J. Chem. Phys., 99, 2695

\bibitem[\protect\citeauthoryear{Wong, Yurchenko, Bernath, Mueller, McConkey
  \& Tennyson}{Wong et~al.}{2017}]{jt686}
Wong A.,  Yurchenko S.~N.,  Bernath P.,  Mueller H. S.~P.,  McConkey S.,
  Tennyson J.,  2017, MNRAS, p. (submitted)

\bibitem[\protect\citeauthoryear{Yadin, Vaness, Conti, Hill, Yurchenko  \&
  Tennyson}{Yadin et~al.}{2012}]{jt529}
Yadin B.,  Vaness T.,  Conti P.,  Hill C.,  Yurchenko S.~N.,   Tennyson J.,
  2012, MNRAS, 425, 34

\bibitem[\protect\citeauthoryear{Yorke, Yurchenko, Lodi  \& Tennyson}{Yorke
  et~al.}{2014}]{jt590}
Yorke L.,  Yurchenko S.~N.,  Lodi L.,   Tennyson J.,  2014, \mn@doi [MNRAS]
  {10.1093/mnras/stu1854}, 445, 1383

\bibitem[\protect\citeauthoryear{Yurchenko \& Tennyson}{Yurchenko \&
  Tennyson}{2014}]{jt564}
Yurchenko S.~N.,  Tennyson J.,  2014, MNRAS, 440, 1649

\bibitem[\protect\citeauthoryear{Yurchenko, Barber  \& Tennyson}{Yurchenko
  et~al.}{2011}]{jt500}
Yurchenko S.~N.,  Barber R.~J.,   Tennyson J.,  2011, \mn@doi [MNRAS]
  {10.1111/j.1365-2966.2011.18261.x}, 413, 1828

\bibitem[\protect\citeauthoryear{Yurchenko, Blissett, Asari, Vasilios, Hill  \&
  Tennyson}{Yurchenko et~al.}{2016}]{jt618}
Yurchenko S.~N.,  Blissett A.,  Asari U.,  Vasilios M.,  Hill C.,   Tennyson
  J.,  2016, \mn@doi [MNRAS] {10.1093/mnras/stv2858}, 456, 4524

\makeatother
\end{thebibliography}

\label{lastpage}

\end{document}